\documentclass[%
 reprint,
superscriptaddress,
 amsmath,amssymb,
 aps,
floatfix,
]{revtex4-2}
\usepackage{}
\usepackage{graphicx}
\usepackage{dcolumn}
\usepackage{bm}
\usepackage{hyperref}

\usepackage[english]{babel}
\renewcommand{\selectlanguage}[1]{}

\begin{document}
\preprint{APS/123-QED}
\bibliographystyle{apsrev4-1}
\title{Data-driven Mean-field within Whole-brain Models}
\author{Martin Breyton}
\email{martin.breyton@univ-amu.fr}
\affiliation{Aix Marseille Univ, APHM, INSERM, INS, Inst Neurosci Syst, Service de Pharmacologie Clinique et Pharmacosurveillance, Marseille, France; Aix Marseille Univ, INSERM, INS}
\affiliation{Aix Marseille Univ, INSERM, INS, Inst Neurosci Syst, Marseille, France}
\author{Viktor Sip}
\author{Marmaduke Woodman}
\author{Meysam Hashemi}
\author{Spase Petkoski}
\author{Viktor Jirsa}
\email{viktor.jirsa@univ-amu.fr}
\affiliation{Aix Marseille Univ, INSERM, INS, Inst Neurosci Syst, Marseille, France}

\date{\today}
\renewcommand\appendixautorefname{Appendix}
\begin{abstract}
Mean-field models provide a link between microscopic neuronal activity and macroscopic brain dynamics. Their derivation depends on simplifying assumptions, such as all-to-all connectivity, limiting their biological realism. To overcome this, we introduce a data-driven framework in which a multi-layer perceptron (MLP) learns the macroscopic dynamics directly from simulations of a network of spiking neurons. The network connection probability serves here as a new parameter, inaccessible to purely analytical treatment, which is validated against ground truth analytical solutions. Through bifurcation analysis on the trained MLP, we demonstrate the existence of new cusp bifurcation that systematically reshapes the system’s phase diagram in a degenerate manner with synaptic coupling. By integrating this data-driven mean-field model into a whole-brain computational framework, we show that it extends beyond the macroscopic emergent dynamics generated by the analytical model. For validation, we use simulation-based inference on synthetic functional magnetic resonance imaging (fMRI) data and demonstrate accurate parameter recovery for the novel mean-field model, while the current state-of-the-art models lead to biased estimates. This work presents a flexible and generic framework for building more realistic whole-brain models, bridging the gap between microscale mechanisms and macroscopic brain recordings.

\end{abstract}

\maketitle


\section{Introduction}
\label{Intro}
Computational models have emerged as essential tools for studying neural activity in neuroscience and translational medicine. Virtual Brain Twins (VBTs) integrate multimodal (structural/functional) data into a coherent mechanistic framework to generate personalized predictions on brain activity \cite{hashemi_principles_2024,breakspear_dynamic_2017,wang_virtual_2024}. This transformative approach has been used recently in clinical trials to provide mechanistic insights into brain dysfunction, and to support personalized diagnosis and treatment planning \cite{jirsa_personalised_2023}. To achieve this, VBTs rely on a whole-brain model, a network of coupled neural models placed at regions and constrained by personalized structural connectivity, to generate individualized large-scale brain activity \cite{sanz-leon_mathematical_2015}. Following source-to-sensor mapping, the generated signals can be compared directly with empirical recordings of brain activity, \textit{e.g.}, measured by electroencephalography (EEG), magneto-encephalography (MEG), or functional magnetic resonance imaging (fMRI). 

A set of plausible mechanisms represented in lumped neural mass models \cite{deco2008dynamic}, is embedded in the parameter space, and the model is inverted using machine-learning (ML) methods \cite{hashemi_simulation-based_2024,breyton_large-scale_2023}. These parameters can be region-specific, reflecting mechanisms taking place at lower scales (neurons or circuits) affecting network properties, for example, the strength of long-range connections between brain regions. In an ideal situation, the parameter space would be directly interpretable as a biological or biophysical quantity (\textit{e.g.} synaptic density, concentration of a neurotransmitter) that could be modulated by an external agent (\textit{e.g.} a pharmaceutical drug). In practice, the interpretation of parameters is shaped by the choice of the neural mass model used to simulate the mesoscopic activity. 

Phenomenological models offer a purely abstract representation of neural activity designed to reproduce key features of brain recordings, such as specific bifurcation \cite{jirsa_nature_2014,saggio_taxonomy_2020} in epilepsy, or oscillatory behavior and functional coactivations between brain regions during resting-state \cite{ponce-alvarez_hopf_2024,xavier_metastable_2024}. Other historical models capture more explicitly some details of brain circuitry \cite{jansen_electroencephalogram_1995,wilson_excitatory_1972}, for example, the interaction between excitatory and inhibitory cells, but they collapse multiple mechanisms into a few effective parameters. Despite their potential \cite{stefanovski_linking_2019}, there remains a mechanistic gap between the parameters of phenomenological models and biological reality that hampers the integration across scales.
In contrast, data-driven ML models can tackle problems with traversing across scales \cite{sip2023characterization}, but lack interpretability, and rely only on patterns in the training data \cite{baker2018mechanistic}. Balancing mechanistic transparency with data-driven flexibility remains a challenging trade-off.

 \textit{Mean-field} derivations of macroscopic activity have introduced analytical solutions for the collective dynamics of neuronal population and interaction thereof. The bottom-up approach of these models maintain the link between some neuronal mechanisms and macroscopic activity, and have shown successful applications \cite{breyton_large-scale_2023,yalcinkaya_personalized_2023,destexhe_computational_2024, jirsa_personalised_2023}. 
Current formalisms, predominantly based on master equations \cite{el_boustani_master_2009} or continuity equations \cite{coombes_next_2023}, rely on a series of restrictive assumptions to ensure analytical closure. These typically include neurons operating under certain constraints, such as limitations to asynchronous irregular spiking,
 \cite{el_boustani_master_2009}, homogeneity \cite{schwalger_towards_2017} or a quenched Lorentzian heterogeneity and thermodynamic limit ($N \rightarrow \infty$) with all-to-all connectivity, \cite{montbrio_macroscopic_2015, coombes_next_2023}. Often, any change in the underlying spiking neuron requires a new derivation \cite{montbrio_macroscopic_2015,Devalle2018,Pietras2025,Gast2021,Gast2023}, and, even slight relaxation in these assumptions can compromise the validity of the resulting mean-field approximation \cite{pyragas_effect_2023,greven_how_2024}. 
Thus, there is a lack of a general framework for deriving \textit{mean-field} models from biologically realistic spiking neuron models that incorporate more degrees of freedom. 
Here, we propose a flexible and accurate method to estimate mean-field dynamics directly from data, and validate it within a whole-brain framework, demonstrating that microscopic phenomena can be inferred from macroscopic brain recordings with greater freedom in the underlying assumptions. To this end, we developed a multilayer perceptron (MLP) network to learn the phase flow of the state variables. We first estimate the macroscopic dynamics, validating the analytically derived approach, and then generalize it to arbitrary probabilities of connections between spiking neurons. Then, we integrate the ML derivation into a whole-brain model, incorporating anatomical data as the network connectivity. Finally, we use simulation-based inference from synthetic and empirical fMRI data to validate the approach at the whole-brain scale.



\section{Ground truth model}
We consider a network of Quadratic-Integrate-and-Fire (QIF; \cite{izhikevich_dynamical_2006}) neurons where the equation of one neuron is given by:
\begin{equation}
\begin{aligned}
\dot{V}_j&=V_j^2 + I_j ,\ \text{if}\ V_j>V_{peak}\ \text{then}\ V_j\leftarrow V_{reset} \\
&\text{with}\ I_j = \eta_j + Js(t) +  I(t)
\label{eq:qif_neuron}
\end{aligned}
\end{equation}
where $I_j$ denotes the input current to the neuron $j$, and $V_{peak}$ is the threshold value that triggers $V$ to be reset to $V_{reset}$. The input current to each neuron is composed of a quenched excitability parameter $\eta_j$, the mean synaptic activation $s(t)$ scaled by the synaptic weight $J$, and $I(t)$ a time-varying component. For a network of infinite size, with an all-to-all connectivity between neurons and a Lorentzian distribution of the excitability parameter $\eta$, the exact macroscopic behavior is given by the so-called firing rate equations \cite{montbrio_macroscopic_2015} (MPR model):
\begin{equation}
\begin{aligned}
&\dot{r}= \Delta/\pi + 2rv \\
&\dot{v} = v^2+\bar{\eta} + Jr +  I(t) -\pi^2r^2
\label{eq:mpr}
\end{aligned}
\end{equation}
where $r$ and $v$ are the mean firing rate and membrane potential, respectively. $J$ is the synaptic weight; and $\bar{\eta}, \Delta$ are the mode and half-width of the Lorentzian distribution of heterogeneity $\eta_j$. 
 
Outside these constraints, the macroscopic behavior does not admit a closed-form solution and remains analytically intractable. Even if simulations remain qualitatively similar (under Gaussian heterogeneity, for example), there are notable differences in the location of the saddle-node bifurcation branches and the cusp point that could be of importance when modeling neurons with non-Lorentzian excitability. Furthermore, some aspects of the QIF network are not captured by the mean-field. For example, the size of the population assumes, in the derivation, the thermodynamic limit, and the connections between neurons are considered to be fully dense, which may not reflect realistic network architectures. Here, we relax the latter assumptions and introduce a connection probability $p$ between pairs of neurons, and show that mean-field dynamics, parameterized by $\bar{\eta}$ and $J$ but also $p$, can be successfully estimated using ML techniques on a finite-size network. 

\section{Multiscale reconstruction}
\subsection{Generating a dataset}
\begin{figure*}
\includegraphics[width=\textwidth]{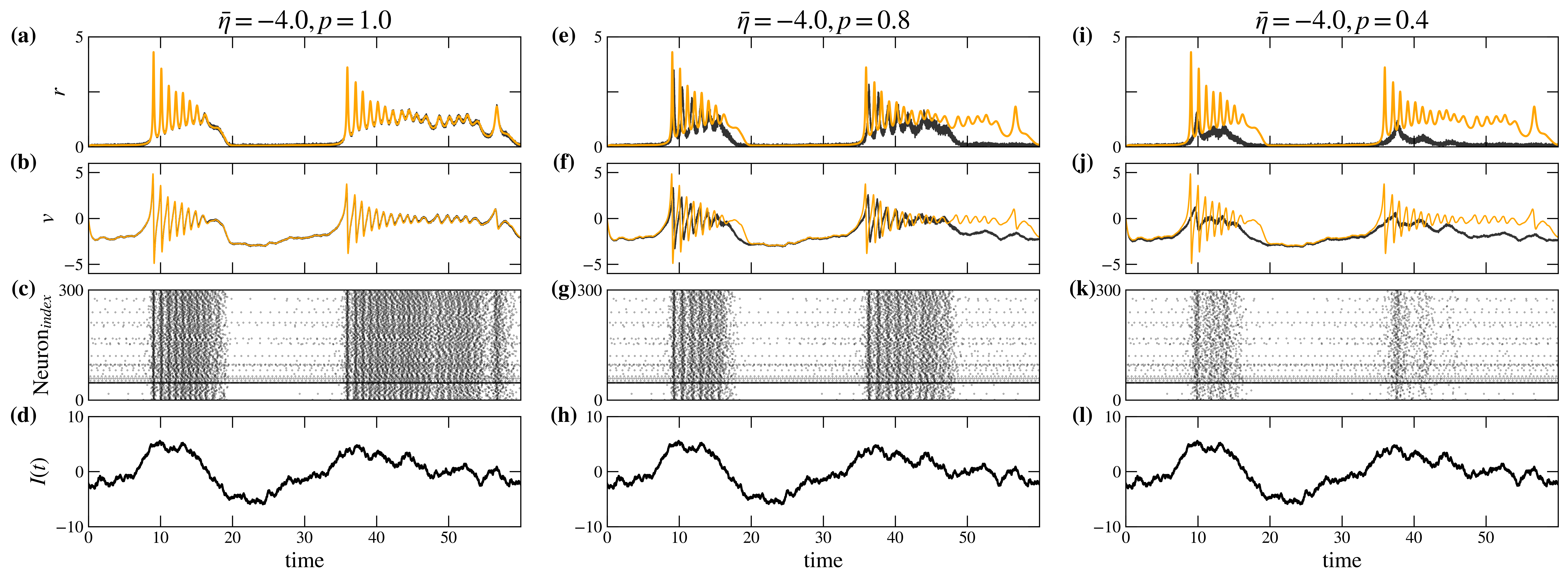}
\caption{\label{fig:data_plot} Numerical simulations of a network of QIF neurons with excitability $\bar{\eta}=-4$ are shown for different levels of connectivity sparsity. Panels (a)-(d) correspond to the all-to-all coupled case with $p=1.0$, (e)-(h) $p=0.8$, and (i)-(l) $p=0.6$. The average firing rate is shown in (a),(e),(i), and average membrane potential in (b),(f),(j), with the QIF network simulations in black and the MPR time series in orange, for the same initial conditions ($r_0=v_0=0$) generated under the same input current $I(t)$ shown in (d),(h),(l). Raster plots of $300$ randomly selected neurons are shown in (c),(g),(k). }
\end{figure*}

Simulations of the QIF network are performed using the Brian2 simulator \cite{Stimberg2019}, with the number of neurons fixed at $N=10^4$. The width of the Lorentzian is fixed at $\Delta=1$, and the synapses are assumed to be instantaneous. Training data is generated by running numerical simulations over a regular grid of parameters ($\bar{\eta}, p,J$). The mode of excitability $\bar{\eta}$ is sampled between $[-8,-1]$ with a step size of $0.05$, the probability of connections between neurons $p$ is homogeneous across neurons and is varied between $[0.6, 1]$ with a step size of $0.1$, and $J$ (constant across neurons) between $[5,15]$ with a step size of $0.5$. Each neuron simultaneously receives identical random input $I(t)$ (\autoref{eq:qif_neuron}) with $1/f$ spectral properties (see Appendix \ref{input.current}), generated using a Fourier transform (see \autoref{fig:data_plot}). Each combination of parameters is run with $5$ different noise seeds for a total of $N=220000$ simulations. For all simulations where $p=1$, the network operates under conditions very close to the mean-field, and MPR times series are also simulated under the same input currents. For each simulation, we extract the average firing rate and membrane potential of the population. Examples of time series for different values of $\bar{\eta}$ and $p$ are shown in  \autoref{fig:data_plot}. As previously shown \cite{montbrio_macroscopic_2015}, when $p=1$, the MPR model almost exactly follows the collective dynamics of the QIF network. However, when $p<1$, we observe discrepancies between the two models, with a damped dynamics for decreasing $p$. For low $p$, even when driven by a strong input current, the collective dynamics fails to exhibit oscillatory behavior, which hints at the absence of the stable focus.

\subsection{Estimating macroscopic dynamics}
There exists a variety of methods for reconstructing dynamical systems from data, ranging from symbolic approaches to methods based on artificial neural networks \citep{DurstewitzEtAl23,YuWang24}. They differ in the model representing the internal dynamics, in the assumptions on the observed variables, the deterministic or stochastic nature of the model dynamics, or the physics- or biology-based constraints on the model form. Here, we use a multilayer perceptron (MLP) \cite{bishop_pattern_2006} to learn the phase flow of the state variables. The objective is to train the neural network to reproduce the deterministic dynamics of the state vector $x$, such that:
\begin{equation}
\dot{x} = \text{MLP}_{\Theta}(x , \{k\},I(t))\ 
\end{equation}
We assume that we fully observe the system under study, such that the state vector is the joint observation of the average firing and membrane potential $x(t)=(r(t),v(t))$, the input drive $I(t)$, and the parameters $\{k\}=\bar{\eta}, J, p$ that constrain the network activity. The set $\Theta=\{W, b\}$ defines the weights and biases of the MLP that we seek to estimate. For training, we downsampled all simulations by a factor $100$, and we split time series into short segments corresponding to $1$ second of QIF activity. Therefore, each training data point is a time series of  $T=100$ discrete time steps, and the $nth$ data point is $\mathbf{x}^n=\{x^n_{t_0}, x^n_{t_1},..., x^n_T, \eta^n, p^n, J^n \} \in \mathbb{R}^{T+3}$.  At each training iteration, only the initial time step is passed on to the neural network. An estimated trajectory is then generated using a Heun integration scheme \cite{kloeden_numerical_1992}, so that $x^n_{t+1} = x^n_{t} + \frac{1}{2}dt (\text{MLP}(x^n_{t}) + \text{MLP}(\tilde{x}^n_{t+1}))$  where the intermediate estimate is $\tilde{x}^n_{t+1}=x^n_t + dt \ \text{MLP}(x^n_t)$. The loss function is defined as the average Euclidean distance between the estimated trajectory and the data: 
\begin{equation}
    \ell^n_{\hat{\Theta}}(\mathbf{\hat{x}}^n,\mathbf{x}^n)=\frac{1}{T} \sum_{t=0}^T\sqrt{(\hat{r}_t^n-r_t^n)^2+(\hat{v}_t^n-v_t^n)^2}
\end{equation}
Training was repeated on MPR and QIF times series data yielding two models denoted as $\text{MLP}_{MPR}(r,v,\bar{\eta},J,I_{ext})$ $\in \mathbb{R}^2$, and $\text{MLP}_{QIF}(r,v,\bar{\eta},J,p,I_{ext})$ $\in \mathbb{R}^2$, respectively. 

\begin{figure*}
\includegraphics[width=\textwidth]{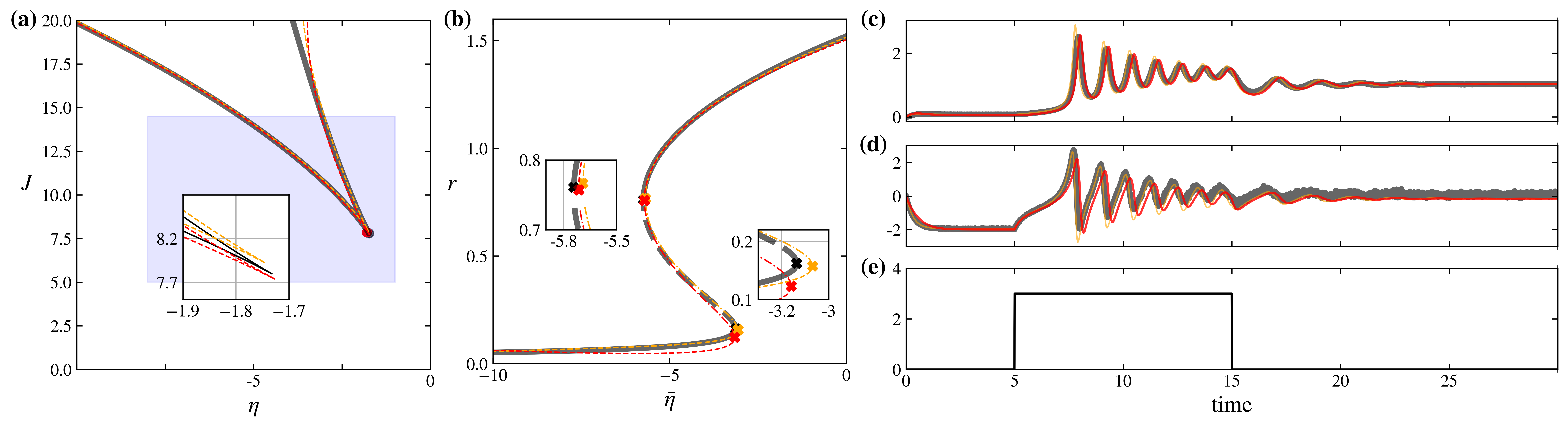}
\caption{\label{fig:mpr_reconstruction} Reconstruction of $\text{MPR}$ dynamics using QIF simulations. (a) Phase diagram comparing the ground truth (thick black line) with the reconstructed dynamics using QIF data (dashed red line) and $\text{MPR}$ times series (dashed orange line). (b) Bifurcation diagram showing ($r,\bar{\eta}$) for $J=15$ and $\Delta=1$. (c) Time series of average firing rate (black) and average membrane potential (green) from a QIF simulation, with $\bar{\eta}=-5, J=15,\Delta=1, p=1$ and initial membrane potential $v_0=0$. (d) Time series of $r(t)$ and $v(t)$ from the MPR model (thick black) and from the reconstructed dynamics for the same initial conditions $r_0=v_0=0$. (e) Input current $I(t)$ used to generate time series in (c) and (d). The cyan shaded area in (a) shows the area in which $\bar{\eta}$ and $J$ were sampled from.}
\end{figure*}

\begin{figure}
\includegraphics[width=\columnwidth]{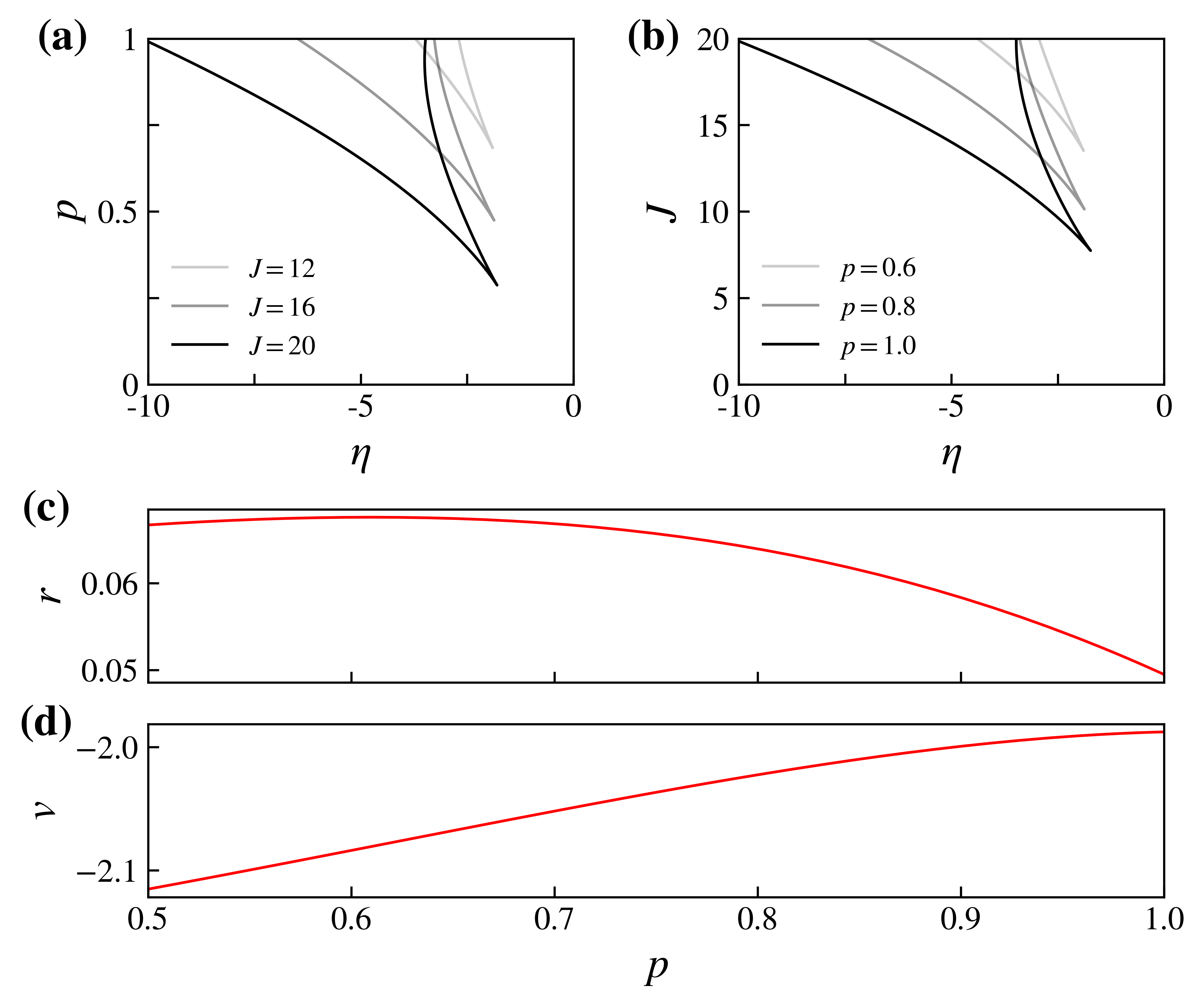}
\caption{\label{fig:pcusp} Phase diagram. (a) Co-dimension two bifurcation (cusp) in $(p, \bar{\eta})$ plane for different values of $J$. (b)  Modification of the cusp in $(J, \bar{\eta})$ plane for different values of $p$ (with $J=15)$. (c) and (d) show the position of the stable fixed point as a function of $p$, obtained using numerical continuation method.}
\end{figure}

After training the MLP, we conduct a bifurcation analysis using a numerical continuation toolkit \cite{jullia_biftoolkit}. We study the following nested function:
\begin{align*}
& \dot{\hat{x}} = \text{MLP}_\mathbf{\hat{\Theta}}(x) = {\hat{W}}^L {a}^{L-1}+ {\hat{b}}^L,\\
& {a}^{\ell} = {\tanh}({\hat{W}}^{\ell}{a}^{\ell-1}+{\hat{b}}^\ell), \\
& \text{ with }\ell \in \{1,...,L-1\}, \ {a}^{0}=x \\
\end{align*}
where $\hat{\Theta}=\{ {\hat{W}}, {\hat{b}} \}$ are the weights and biases frozen after training. We present the results of a bifurcation analysis in \autoref{fig:mpr_reconstruction} for $\text{MLP}_{MPR}$ and $\text{MLP}_{QIF}|_{p=1}$. The results show that the phase diagram is almost perfectly retrieved by the MLP in both cases. The chaotic regime found under a periodic input was also recovered (see Appendix \ref{chaos.valid}). We find that the retrieval of the bifurcation points is slightly more precise when the reconstruction is based on the MPR times series. Moreover, we find that the branches of the cusp are correctly extrapolated outside the training space. The estimation of mean-field dynamics based on QIF data now allows us to study the bifurcation topology with respect to the parameter $p$ (\autoref{fig:pcusp}). There is a cusp bifurcation in the co-dimension spanned by $\bar{\eta}$ and $p$, very similar to the one that exists between $\bar{\eta}$ and $J$. The relationship between $p$ and $J$ is almost symmetric, as decreasing $p$ for a fixed $J$ has the same effect on the cusp as decreasing $J$ for a fixed $p$. We also verify that there is no bifurcation along the $p$-direction by showing the coordinates $r$ (\autoref{fig:pcusp}(c)) and $v$ (\autoref{fig:pcusp}(d) of the stable fixed point as a function of $p$.

\subsection{Whole-brain model}
\begin{figure*}
\includegraphics[width=\textwidth]{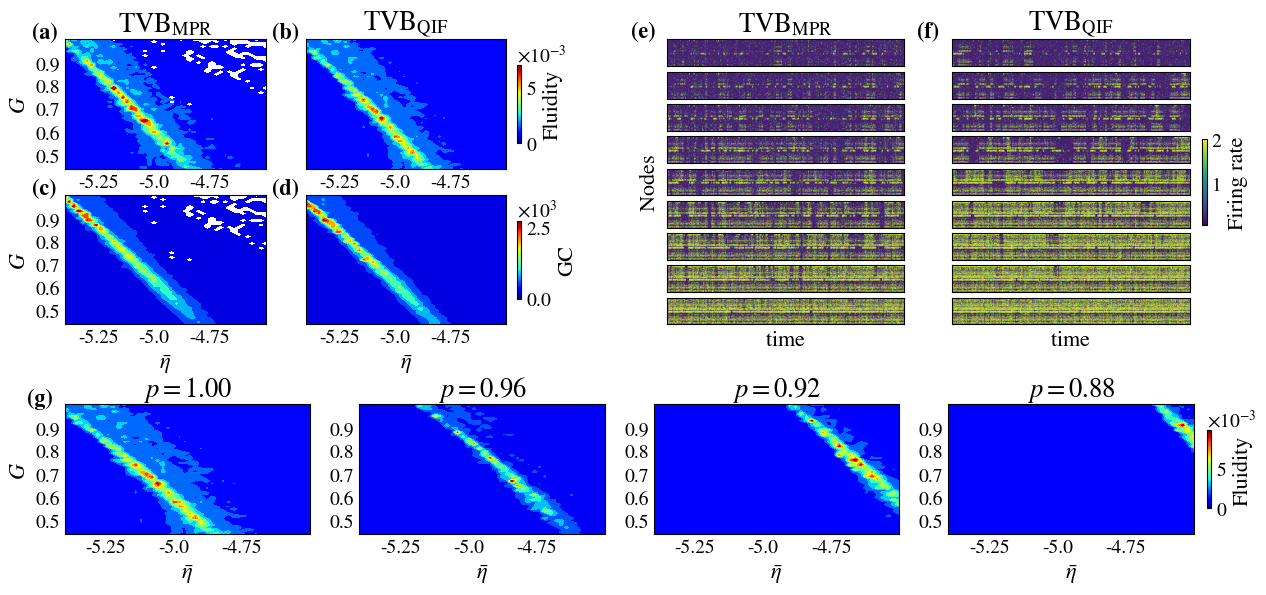}

\caption{\label{fig:TVB_comp} Comparison of the dynamics of whole-brain models using the ground truth model ($\text{MPR}$) or the reconstructed model $\text{MLP}_{\text{QIF}}$ as a neural mass. (a),(b) Fluidity of the dynamics in the ($G,\bar{\eta}$) space for $\text{TVB}_{\text{MPR}}$ and $\text{TVB}_{\text{QIF}}$. (c),(d) Global connectivity in the ($G,\bar{\eta}$) space for $\text{TVB}_{\text{MPR}}$ and $\text{TVB}_{\text{QIF}}$. (e),(f) carpet plots of \textit{firing rate} activity for increasing values of $G$ (ranging from $0.58$ to $0.66$ from top to bottom), with $\bar{\eta}=-5$ and $J=15$. (g) Fluidity of the dynamics shown for different probability of connections between neurons, from left to right $p=1$ (same as (b)), $p=0.96$, $p=0.92$$p=0.88$ simulated with simulated with and $\text{TVB}_{\text{QIF}}$ and $J=15$. The white spots in (a) and (c) correspond to missing values.
}
\end{figure*}
An important application of mean-field models of spiking neurons is their integration into whole-brain models of neural activity. In this context, mean-field models are often recast as neural mass models that simulate the activity of a small patch of brain tissue. They are then coupled together to build a network of interacting brain regions, allowing simulation of large-scale brain dynamics. The standard model for brain network modeling is defined as \cite{Jirsa1996, Jirsa2000}:
\begin{equation}
    \dot{\psi_i}=\mathcal{N}(\psi_i,\{k\})+g \sum_{j}W_{ij}S(\psi_j(t-\tau_{ij})) + \xi_i(t)
\label{eq:tvb}
\end{equation}
where $\psi_i$ is the field activity of node $i$ in the network at time $t$, and $\dot{\psi_i}$ is its derivative w.r.t time. $\mathcal{N}$ is a function (usually a nonlinear differential equation) that governs the dynamics of each node. It is a function of the current state $\psi$ and a set of parameters $k$ that depends on the choice of neural mass model, and can encompass abstract quantities such as excitability, or it can refer to more concrete physiological parameters like synaptic strength. The matrix $W$ is the structural connectivity between nodes of the network. It is estimated from diffusion weighted imaging by reconstructing the subject-specific connectome \cite{sporns_human_2005}, which traces the white fiber tracts in the brain. $S$ is a synaptic activation function, which here is set to be the identity function, and $\xi_i$ is some dynamical noise specific to $i-$region. The coupling term is scaled by a global coupling parameter $g$ \cite{rabuffo2025mapping}. In addition, we set up an observation function to model signals measured by functional magnetic resonance imaging (fMRI). Specifically, we use a haemodynamic model \cite{friston_nonlinear_2000} to transform the source activity to a Blood Oxygenation Level Dependent (BOLD) signal. 

We compare whole-brain simulations using two different neural mass models. First, the ground truth Montbrió et al. model denoted as $\mathcal{N}(x,\{k\})=\text{MPR}(x,\{k\})=(\dot{r},\dot{v})$ (see \autoref{eq:mpr}), and second, the result of mean-field estimation trained on QIF time series $\mathcal{N}(x,\{k\},I(t))=\text{MLP}_{\text{QIF}}(x,\{k\},I(t))$ (from Appendix \ref{lin.coup}). We refer to the corresponding whole-brain model as $\text{TVB}_{\text{MPR}}$ and $\text{TVB}_{\text{QIF}}$ models. We set the parameters $k=\bar{\eta}$ to be homogeneous across brain regions, we fix the synaptic term $J=15$. We perform parameter sweeps over $\bar{\eta}$ and global coupling $g$ using the two different models. The activity of the whole-brain is characterized by the functional connectivity (FC) between regions and its dynamics, the functional connectivity dynamics (FCD), which is obtained by sliding a window of $FC$ matrices \cite{hansen2015functional}. We measure the \textit{fluidity} of the system's emergent dynamics by the variance of the upper triangular part of the FCD, denoted as $\text{Var}(FCD_{Utri})$, and the global connectivity (GC) by summing all entries of the FC matrix. In \autoref{fig:TVB_comp}, we present the fluidity of the dynamics in the parameter space for the two different models.

We find that there is a ridge in the parameter space where the model can produce fluid dynamics and coherent activity in terms of global connectivity. This ridge is distributed similarly in the ($G,\bar{\eta}$) parameter space in both the $\text{TVB}_{\text{MPR}}$ and $\text{TVB}_{\text{QIF}}$ models. The trade-off between global connectivity and fluidity is also preserved in the reconstructed model. This is further visible by direct inspection of the time series for different values of $G$ (and $\bar{\eta}=-5$). The time series confirm that the transition between the low-activity regime (\autoref{fig:TVB_comp} (e),(f) Top) and the high-activity regime where most nodes are in the upstate (\autoref{fig:TVB_comp} (e),(f) Bottom) through a non-trivial regime with cascades of co-activation is both quantitatively and qualitatively similar in both models. The white spots in \autoref{fig:TVB_comp} (a) and (c) correspond to missing values due to numerical instabilities in the simulations.

Now that $p$ is a parameter of the neural mass model, we can explore (\autoref{fig:TVB_comp}) how changing the connectivity between neurons affects the whole-brain dynamics. The ridge of non-trivial dynamics is progressively shifted as $p$ decreases in a direction perpendicular to the ridge itself. This corresponds to the overall damping effect of $p$ on the activity and the need to compensate by higher excitability or coupling to reach the same level of fluidity.

\subsection{Model inversion}
One of the key goals of whole-brain modeling is to estimate parameter values from empirical imaging data to unveil the underlying mechanisms and to inform and improve targeted interventions \cite{hashemi_principles_2024}. This process of model inversion--finding the best set of parameters-- can be done through optimization techniques (least squares, gradient descent; \cite{deco2021dynamical,kong2021sensory}) or Bayesian inference \cite{hashemi2020bayesian, jha2022fully}. The latter, typically implemented via Markov chain Monte Carlo (MCMC) sampling offers a full characterization of the posterior distributions, which is critical for identifying degeneracy between parameters \cite{van_de_schoot_bayesian_2021}. To address the computational and convergence issues of this approach, here, we use the \textit{simulation-based inference} (SBI) framework \cite{cranmer2020frontier}. SBI is a class of likelihood-free and amortized inference methods that relies on artificial neural networks \cite{papamakarios_masked_2018} to estimate an invertible mapping between the parameter space of the model and low-dimensional feature space of choice. To do this, first, a set of random simulations is run for parameter values $\theta$ sampled from prior distributions. Second, a set of features $x$ is extracted from each simulation, here based on FC and FCD matrices (see Appendix \ref{sbi}). Third, a neural network called \textit{deep density estimator} \citep{gonccalves2020training, liu2021density, hashemi_amortized_2023} is trained to estimate the amortized posterior distribution $\mathcal{P}(\theta|x)$. The final step is to provide the feature vector $x_0$ extracted from an observation, and compute the posterior distributions $\mathcal{P}(\theta|x_0)$. One common way to estimate the reliability of the inference on simulated data is to compute, for each observation, the posterior $\textit{z-score}$, $z_n=\Big|\frac{\mu_n-\theta_n}{\sigma_n}\Big|$ and the posterior shrinkage $s_n=1-\frac{\sigma_n^2}{\tau_n^2}$ where $\theta_n$ is a component $n$ of the ground truth vector drawn from the prior, $\mu_n$ the posterior mean, $\sigma_n$ the posterior standard deviation and $\tau_n$ the prior standard deviation \cite{betancourt2018calibratingmodelbasedinferencesdecisions}. Together, they evaluate the distance between the (mean) inferred value and the actual ground truth value, and if the variance of the posterior distribution has decreased compared to the prior.
\begin{figure}
\includegraphics[width=\columnwidth]{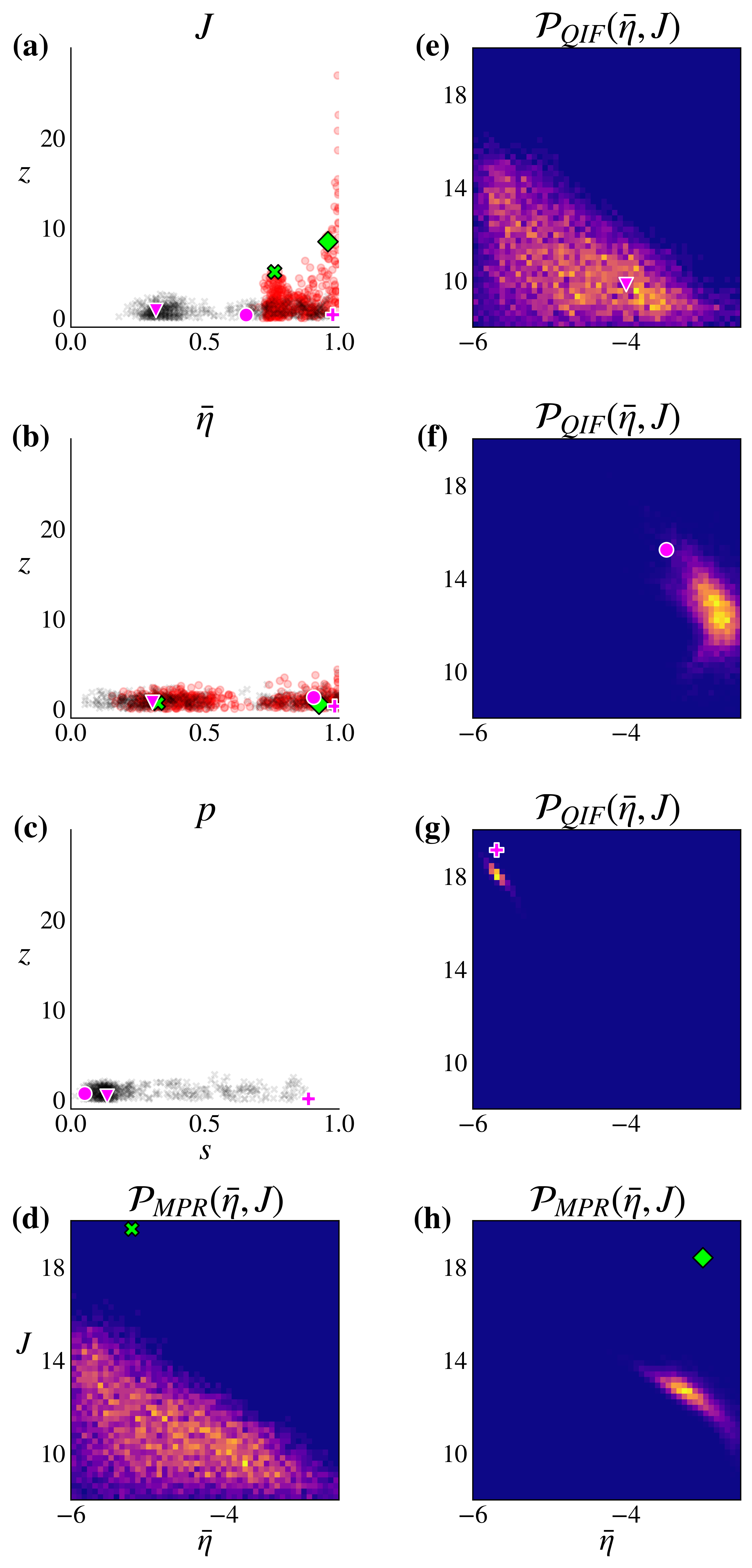}
\caption{\label{fig:infval} Posterior \textit{z-score} ($z$) and shrinkage ($s$) for parameters $J$, $\bar{\eta}$ and $p$. In (a)(b)(c) the black points are inferred using the $\text{TVB}_{\text{QIF}}$ model and the red points using the $\text{TVB}_{\text{MPR}}$ model. The colored symbols correspond to specific manually selected points for which the joint posterior distributions of $\bar{\eta}$ and $J$ are shown in (d) to (h), along with the true values of $\bar{\eta}$ and $J$. In (d),(h) and (e),(f),(h) $\mathcal{P}_{MPR}(\bar{\eta},J)$ and $\mathcal{P}_{QIF}(\bar{\eta},J)$ are the joint posterior distributions of $\bar{\eta},J$ resulting from the inference using the $\text{TVB}_{\text{MPR}}$ and $\text{TVB}_{\text{QIF}}$ model respectively. All ground truth values were generated using the  $\text{TVB}_{\text{QIF}}$ model, with different values of $p$ and $J=15$.
}
\end{figure}

This procedure was repeated for the whole-brain models $\text{TVB}_{\text{MPR}}$ and $\text{TVB}_{\text{QIF}}$. In all cases, we assigned a homogeneous value for the parameters $\bar{\eta},J$ and $p$ across brain regions and fixed the global coupling and noise intensity to reasonable values, previously identified, $g=0.55,\sigma=0.34$ \cite{breyton_large-scale_2023}. Parameters $\bar{\eta}$, $J$ and $p$, were drawn from uniform prior distributions: $\bar{\eta}\sim \mathcal{U}([-6, -2.5])$, $J \sim \mathcal{U}([8, 20)$ and $p \sim \mathcal{U}([0.6, 1])$. The boundaries of the priors were chosen wide enough so that all dynamical regimes from the neural mass model could be expressed. After training, we obtained two different posterior distributions, one corresponding to each $\mathcal{P}_{MPR}, \mathcal{P}_{QIF}$. 

We then generated 500 ground truth observations using $\text{TVB}_{\text{QIF}}$ with different combinations of $\{\bar{\eta}, J,p\}$ drawn from the prior distribution, and in \autoref{fig:infval}, we display the posterior $\textit{z-scores}$ and shrinkage $\mathcal{P}_{MPR}$ (red points) and $\mathcal{P}_{QIF}$ (black points) for all of them. All posterior distributions $\mathcal{P}_{QIF}$ have a low $\textit{z-score}$ value which indicates that the estimation is always close to, or contains the ground truth. But, the shrinkage, which reflects the level of `confidence' of the estimation has a lot of variability. This is illustrated in \autoref{fig:infval} (e),(f),(g) where we show the position of the ground truth (purple symbols) and the shape of the posterior distribution in the ($\bar{\eta},J$) space. On the contrary, the posterior distributions $\mathcal{P}_{MPR}$, all show good shrinkage in the estimation of $J$ since all points in \autoref{fig:infval} (a) have $s>0.5$. The \textit{z-score} values, however, are much higher that for $\mathcal{P}_{QIF}$. As illustrated in \autoref{fig:infval} (d),(h), the green symbols are far from a narrow posterior density, meaning that the estimations are `confidently wrong'.

\begin{figure}
\includegraphics[width=\columnwidth]{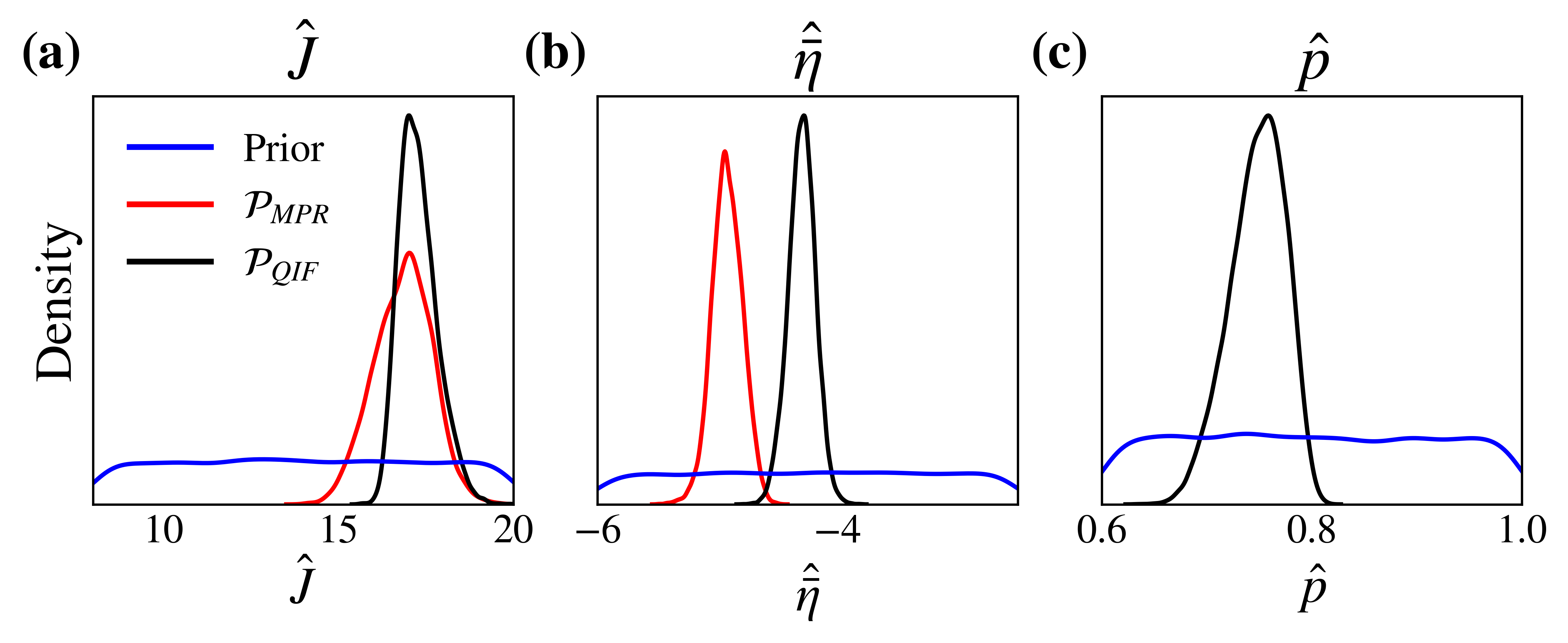}
\caption{
\label{fig:infdata}
Posterior distributions estimated from the fMRI data of a real subject using the same data features as in \autoref{fig:infval} (see Appendix \ref{sbi} for details). Marginal posterior distributions: (a) posterior distribution of $\hat{J}$ (b) posterior distribution of $\hat{\eta}$ (c) posterior distribution of $\hat{p}$. In blue is the prior distribution for each parameter. In red are the posterior distributions obtained using the $\text{TVB}_{\text{MPR}}$ model. And in black the posterior distributions obtained using the $\text{TVB}_{\text{QIF}}$ model.
}
\end{figure}
Lastly, we used the fMRI data from the subject whose DTI data were used to build the connectome and run the simulations. After extracting the same features from the real BOLD data, we analyze the inference results using different models (\autoref{fig:infdata}). We find that the two models converge to two different solutions. They agree on the posterior distribution of synaptic strength $J$, but the value of excitability $\bar{\eta}$ is underestimated by the $\text{TVB}_{\text{MPR}}$ model. According to the $\text{TVB}_{\text{QIF}}$ model, excitability is higher but the connectivity between neurons is not all-to-all ($p<0.8$).

\section{Conclusion}
We provide a flexible method for building a whole-brain model of neuronal activity directly from spiking neurons using machine learning. We showed that this method can accurately estimate and reproduce the dynamical properties of a QIF network, as previously identified analytically. The approach goes beyond traditional mean-field derivation, as it is not limited in the choice of parameterization to be considered. To highlight this, an `augmented' Montbrió et al. model \cite{montbrio_macroscopic_2015} can be constructed by including the probability of connection between neurons $p$ in its parameterization. A bifurcation analysis with respect to this additional parameter was presented. We also demonstrated that an artificial neural network can be integrated into a whole-brain modeling framework and behaves like a traditional neural mass model. 

The spiking network model that we considered (and the corresponding \textit{mean-field}) is in itself a minimalist representation of neuronal dynamics, and the new parameter $p$ we introduced is highly degenerate with $J$. This choice was motivated by the need of a ground truth model with known applications in the field of whole-brain modeling \cite{Lavanga2023,breyton_large-scale_2023,Fousek2024,Rabuffo2021}. Furthermore, even for this spiking model the effect of connection probability on mean-field dynamics is a subject of theoretical interest  \cite{greven_how_2024} and biological relevance \cite{ko_functional_2011}. This shows the relevance of our method, where any parameter accessible in the microscale simulator can be used in the parameterization. Despite the obvious degeneracy between $p$ and $J$, our results on synthetic data show that if the true underlying process is the result of a joint effect of the two mechanisms, inferring using an incompletely parameterized model often leads to biased estimations. Without entering into the interpretation of the estimated parameters, this is further emphasized by our result on real fMRI data where two different solutions are found from the same data features. This reveals that the same observation (\textit{i.e.,} the same data features) can be explained by different parameter combinations, highlighting the challenge of non-identifiability in model-based inference \cite{dangelo_quest_2022, hashemi_simulation-based_2024}.

One limitation of the method is the need to consider a fully observed system. Other dynamical system reconstruction methods focus on partially observed systems and reconstruction of a latent phase space of unknown dimension \cite{sip2023characterization}. While these methods are very powerful, they are more suited for truly unknown systems, such as empirical recordings of neuronal activity. Here, we developed a simpler method that can be applied in controlled experimental setups, such as simulations. The advantage is that the result of the estimation dynamics can be readily integrated into a whole-brain modeling framework. 
Also, we restricted our analysis to a simple network of spiking neurons that is missing many features that can be found in other models (such as synaptic activation and adaptation \cite{chen_exact_2022}, or neuronal heterogeneity \cite{Gast2023}). This choice was motivated by the necessity to compare our results to a model avaiable in an analytical closed form with known applications at the whole-brain level \cite{Lavanga2023,breyton_large-scale_2023,Fousek2024,Rabuffo2021,hashemi_simulation-based_2024}.

In sum, the method presented here offers a flexible approach for studying the effect of microscopic parameters at the whole-brain level. Specifically, the method could be used with heavily detailed simulators of neural networks \cite{Markram2015,Yao2022,Romani2024,Hjorth2020}
for which the mean-field derivations are intractable. In these models, subtle biological mechanisms can be systematically explored \cite{GuetMcCreight2024,Carannante2024}, and could potentially be parameterized in a data-driven macroscopic model. This could be particularly critical for application of VBTs in disorders such as psychiatry, where most treatments are pharmacological and involve complex interactions between receptor activation and physiological responses. 

\begin{acknowledgments}
This project/research has received funding from the European Union’s Horizon Europe Programme under the Specific Grant Agreement No. 101137289 (Virtual Brain Twin Project), No. 101147319 (EBRAINS 2.0 Project), and No. 101057429 (project environMENTAL).
\end{acknowledgments}

\appendix
\section{Chaotic behavior}\label{chaos.valid}
\begin{figure}
\includegraphics[width=\columnwidth]{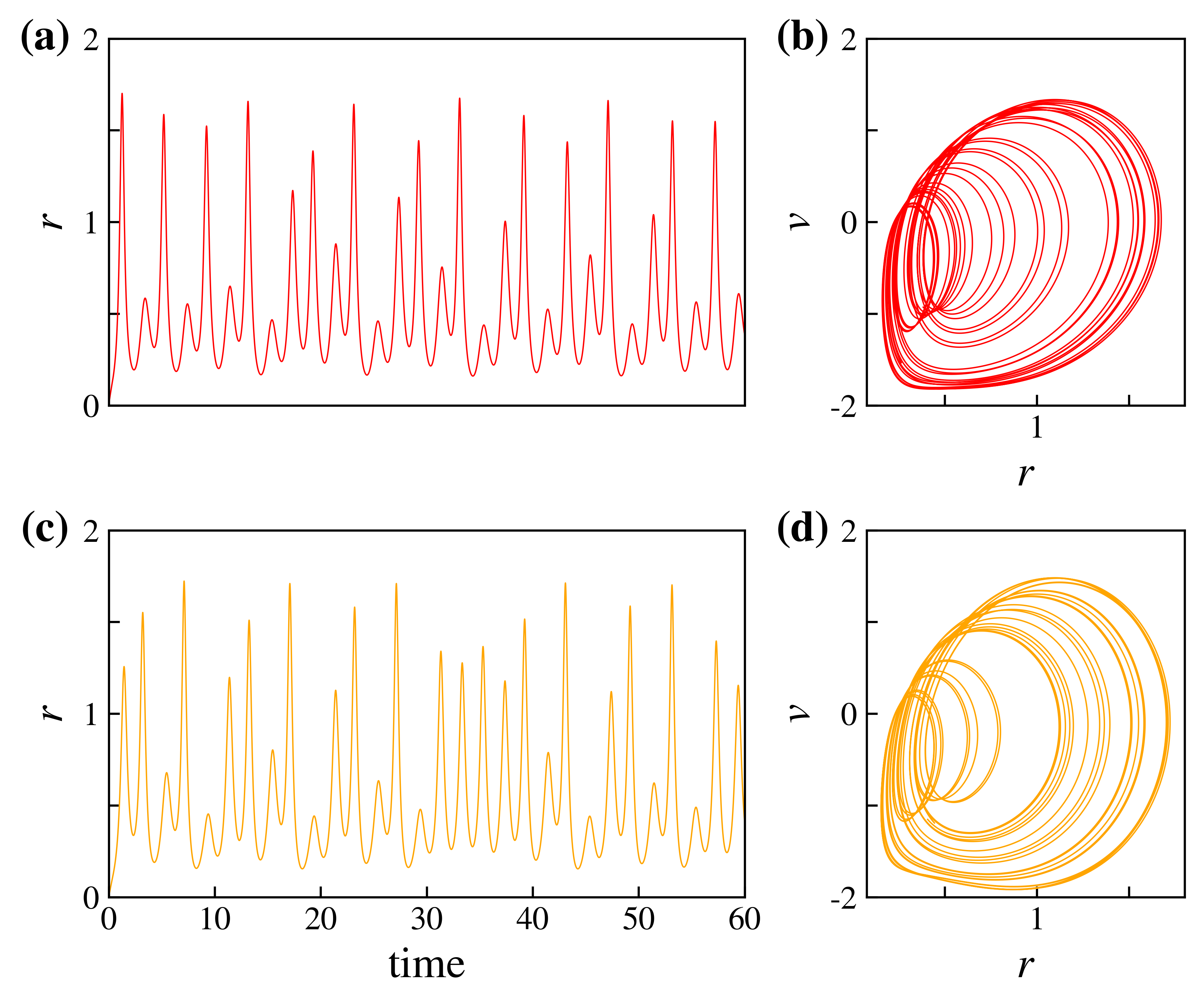}
\caption{\label{fig:chaos} The chaotic regime found in the analytical MPR model is also retrieved by the MLP estimation. Panels (a) and (b) show the firing rate ($r$) times series, and corresponding trajectories in the phase-space of ($r$, $v$) of the $\text{MPR}$ model.  Panels (c) and (d) show the same using $\text{MLP}_{QIF}$ estimation. Time series generated under a sinusoidal input current with amplitude $I_0=3$ and frequency $\pi$, and parameters $\bar{\eta}=-2.5$ and $J=10.5$.} 
\end{figure}

In \cite{montbrio_macroscopic_2015}, a chaotic regime was identified for a specific range of parameters. In \autoref{fig:chaos}, we reproduced this regime for $\bar{\eta}=-2.5$ and $J=10.5$ under a sinusoidal input of amplitude $I_0=3$ and frequency $\pi$. This chaotic regime was not explicitly sampled in the training data but emerged from the reconstructed dynamics.

\section{Linear coupling}\label{lin.coup}
The whole-brain model was constructed by introducing linear coupling between the neural mass models in \autoref{eq:tvb}. For this to be valid it is required that:
\begin{equation}
    \dot{\psi_i}=\mathcal{N}(\psi_i,I(t),\{k\})=\mathcal{N}(\psi_i,\{k\})+I(t).
\end{equation}
This holds true for the MPR model, since $I(t)$ enters linearly in the $\dot{v}$ equation (see \autoref{eq:mpr}), so that if we write:
\begin{align*}
    f(r, v, I(t)) &= (f_1, f_2) \\
    &= (\Delta/\pi + 2rv, v^2+\bar{\eta}+Jr+I(t)-(\pi r)^2).
\end{align*}
We have $\frac{\partial f_1}{\partial I_{ext}}=0$ and $\frac{\partial f_2}{\partial I_{ext}}=1$. We therefore build the whole-brain model with $I(t)=g \sum_{j}W_{ij}S(\psi_j(t-\tau_{ij}))$. To check whether this is valid for the reconstructed models, we check that the MLP can be linearized with respect to $I_{ext}$. We leverage the automatic differentiation framework \cite{jax2018github} for training to study the following partial derivatives $\frac{\partial \hat{\dot{r}}}{\partial I_{ext}}$ and $\frac{\partial \hat{\dot{v}}}{\partial I_{ext}}$ considering $\mathbf{MLP_{\Theta}}=(\hat{\dot{r}},\hat{\dot{v}})$. We find that the partial derivatives are not exactly constant in the $(r,v)$ space:
\begin{align*}
&\dot{\hat{\boldsymbol{x}}} = \mathbf{MLP}_{\hat{\Theta}}(\boldsymbol{x} , \eta, p, I_{ext}) \neq \mathbf{MLP}_{\hat{\Theta}}(\boldsymbol{x} , \eta, p,\mathbf{I_{ext}}=0)+\mathbf{I_{ext}}
\end{align*}
Even though coupling linearity cannot be truly assumed, we find that when enforcing it, the behavior of the model still remains consistent.  
\begin{figure}
\includegraphics[width=\columnwidth]{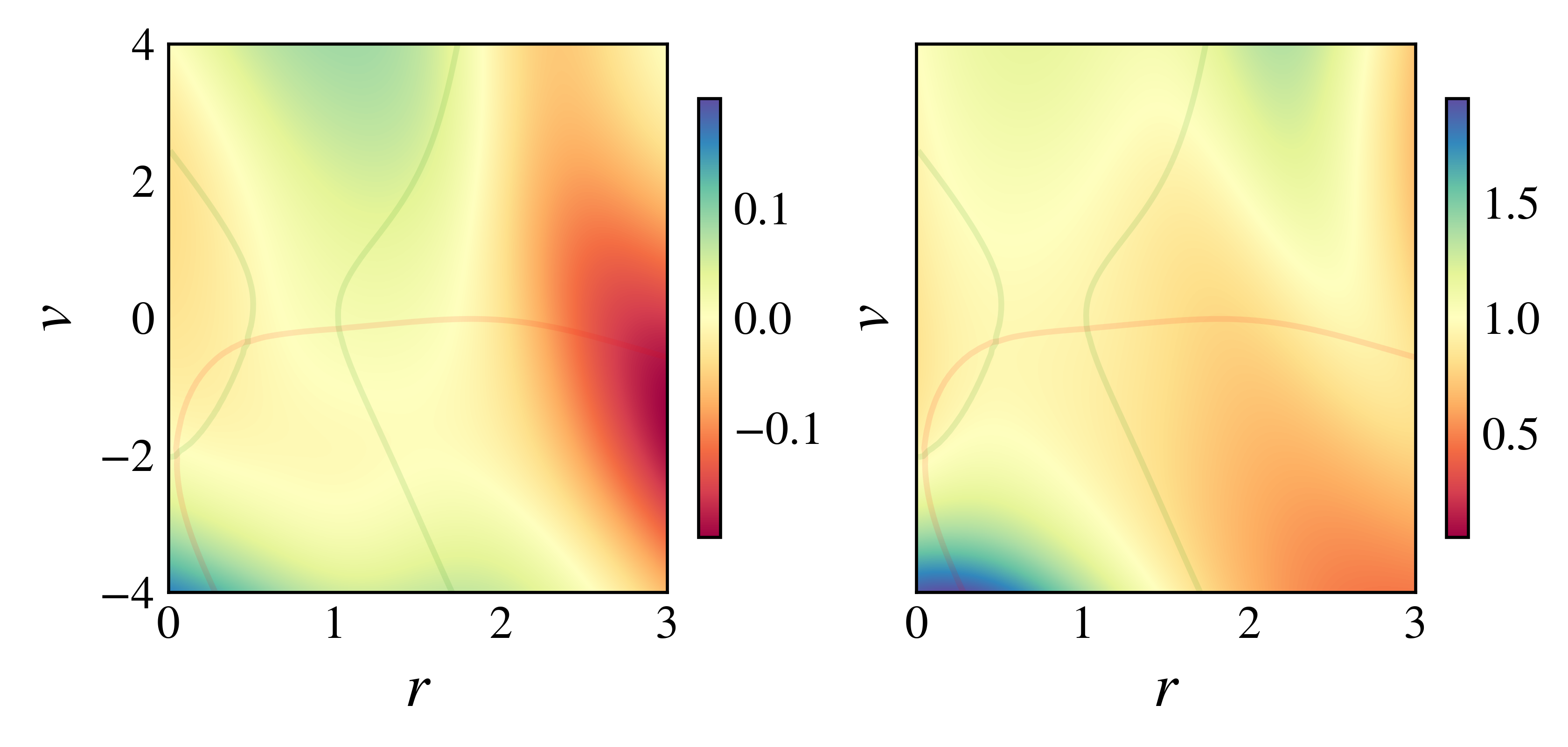}
\caption{\label{fig:jacobian} Values of the partial derivatives on the phase space of the reconstructed dynamics. $\partial \hat{\dot{r}} / \partial I_{ext}$ on the left and $\partial \hat{\dot{v}} / \partial I_{ext}$ on the right. The shaded red and green lines are the nullclines of $r$ and $v$ for the $\text{MLP}_{QIF}$ model, respectively, obtained using a marching algorithm. The values of parameters are set as $J=15$, $p=1$, and $I_{ext}=0$.}
\end{figure}

\section{Input current}\label{input.current}
\begin{figure}
\includegraphics[width=\columnwidth]{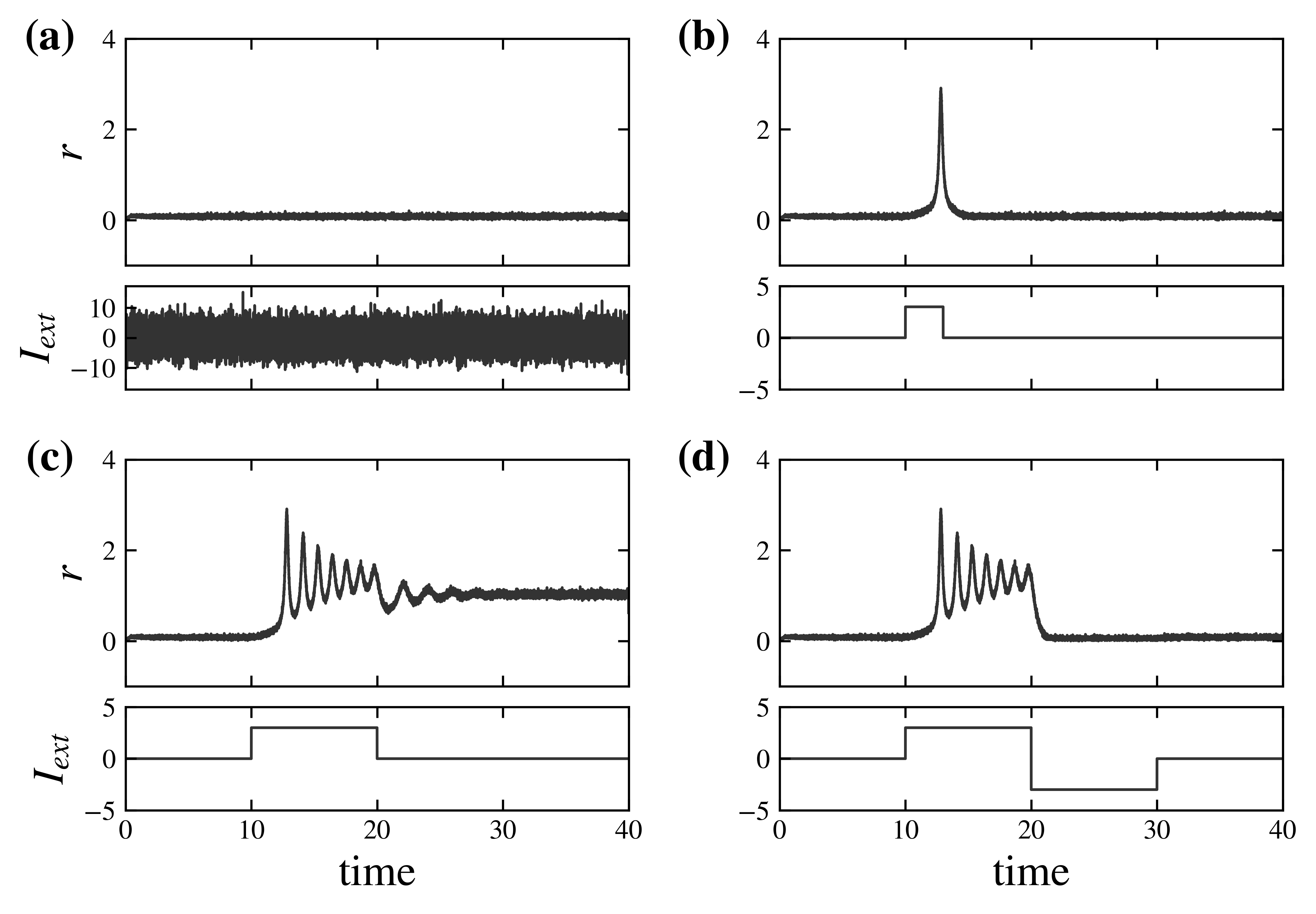}
\caption{\label{fig:input_current} Firing rate times series of QIF simulations under different input currents. Given the same parameters $\bar{\eta}=-5, J=15, p=1$, the transition between the stable fixed point and the stable focus depends on the input current. In (a) a gaussian noise ($\mu=0, std=3)$, in (b) a pulse of amplitude $3$ with a duration of $3$, in (c) the same amplitude with a duration $10$, and in (d) a positive pulse of duration $10$ followed by a negative pulse of the same duration.}
\end{figure}

One key aspect of the method we presented is to generate a dataset that is sufficiently expressive of the dynamics under study. In other words, the phase space must be sufficiently explored in order to reconstruct attractors and bifurcations. We considered three approaches to sample the low-dimensional phase space: noise exploration, initial condition sampling, and stimulation. When simulating a network of QIF neurons, the initial average value of the membrane potential can be easily manipulated by setting the initial potential of each neuron. However, the firing rate is an emerging property of the network and initializing it at different values would require clever engineering, if it is even possible. In \autoref{fig:input_current}, we highlight the evolution of the collective firing rate under different input currents. A purely noise-driven approach using a Gaussian noise, does not trigger the collective dynamics to switch to the stable focus even when it exists ($\bar{\eta}=-5, J=15$). When stimulated by a constant step current of amplitude $I_0=3$, we find that if the duration of the stimulus is sufficient (here shown for $10s$), the firing rate moves away from the stable fixed point towards the stable focus and stays there after release. If the same stimulus is applied for a shorter time, the firing rate does make an excursion towards the stable focus but returns the stable fixed point after release. The approach we chose is shown in \autoref{fig:data_plot}. We constructed a $1/f$ noise by using a Fourier transform:
\begin{align}
x_{\text{white}}(n) &\sim \mathcal{N}(0, \sigma^2) \\
X_{\text{white}}(k) &= \sum_{n=0}^{N-1} x_{\text{white}}(n) \, e^{-i 2\pi \frac{kn}{N}} \\
S(k) &=  \left( \frac{1}{\left(\frac{k}{N}\right)^{\text{exponent}}} \right), k > 0 \\
S_{\text{norm}}(k) &= \dfrac{S(k)}{\sqrt{\dfrac{1}{K} \sum_{j=0}^{K-1} S(j)^2}} \\
X_{\text{1/f}}(k) &= X_{\text{white}}(k) \times S_{\text{norm}}(k) \\
x_{\text{1/f}}(n) &= \dfrac{1}{N} \sum_{k=0}^{\lfloor N/2 \rfloor} X_{\text{1/f}}(k) \, e^{i 2\pi \frac{kn}{N}} + \text{(mirror terms)}
\end{align}

\section{Data features}\label{sbi}
Functional Connectivity (FC) and Functional Connectivity Dynamics (FCD) are widely used to characterize the statistical dependencies and temporal variability of brain activity across distinct regions. FC is typically defined as the Pearson correlation matrix computed from multivariate time series data. Given regional BOLD signals arranged in a matrix  $\mathbf{X} \in \mathbb{R}^{N  \times T}$, where $N$ is the number of brain regions and $T$ the number of time points, the static FC matrix is computed as $\mathbf{FC} = \text{corr}(\mathbf{X})$. Each element $\mathbf{FC}_{ij}$ represents the linear correlation between the time series of regions $i$ and $j$, resulting in a symmetric matrix with unit diagonal. To capture the temporal dynamics of functional interactions, a sliding-window approach is employed to construct the FCD matrix. The time series $\mathbf{X}$ is segmented into overlapping windows of fixed length $L$ with a step size $S$. For each window $w$, a corresponding FC matrix $\mathbf{FC}^{(w)}$ is computed as above. These matrices are then vectorized by extracting their upper triangular elements (excluding the diagonal), forming a sequence of vectors ${\mathbf{f}^{(1)}, \dots, \mathbf{f}^{(W)}}$. The FCD matrix is constructed by computing the pairwise Pearson correlation between these vectors: $\mathbf{FCD}_{ij} = \text{corr}(\mathbf{f}^{(i)}, \mathbf{f}^{(j)})$, resulting in a symmetric matrix of size $W \times W$, where $W$ is the total number of windows. For downstream inference tasks, each FC or FCD matrix $\mathbf{A} \in \mathbb{R}^{n \times n}$ is summarized using a set of statistical summary. The upper triangular entries (excluding the diagonal), denoted $\mathbf{A}_{ut} = \{ {A}{ij} \mid i < j \}$, are used to compute a collection of descriptive statistics: sum, mean, standard deviation, minimum, maximum, skewness, and kurtosis. In addition, quantiles (5th, 25th, 50th, 75th, and 95th percentiles) are computed from the full matrix $\mathbf{A}$ to capture the distributional shape.
Spectral properties of the matrix are characterized by performing eigenvalue decomposition $\mathbf{A} = \mathbf{V} \boldsymbol{\Lambda} \mathbf{V}^{-1}$, followed by the application of the same descriptive statistics to the real parts of the eigenvalues ${\lambda_1, \dots, \lambda_n}$. To further capture low-dimensional patterns, Principal Component Analysis (PCA) is applied to $\mathbf{A}$, retaining a fixed number of components (e.g., 3), and computing descriptive statistics over the projected components.
An additional global measure is obtained by computing the absolute off-diagonal sum $S_{\text{off}} = \sum_{i \neq j} |A_{ij}|$.
Together, these features provide a compact yet comprehensive summary of the structural and statistical characteristics of both static and dynamic functional connectivity matrices, facilitating their use in predictive modeling and statistical inference.

\section{Training benchmark}\label{train.bench}
\begin{figure}
\includegraphics[width=\columnwidth]{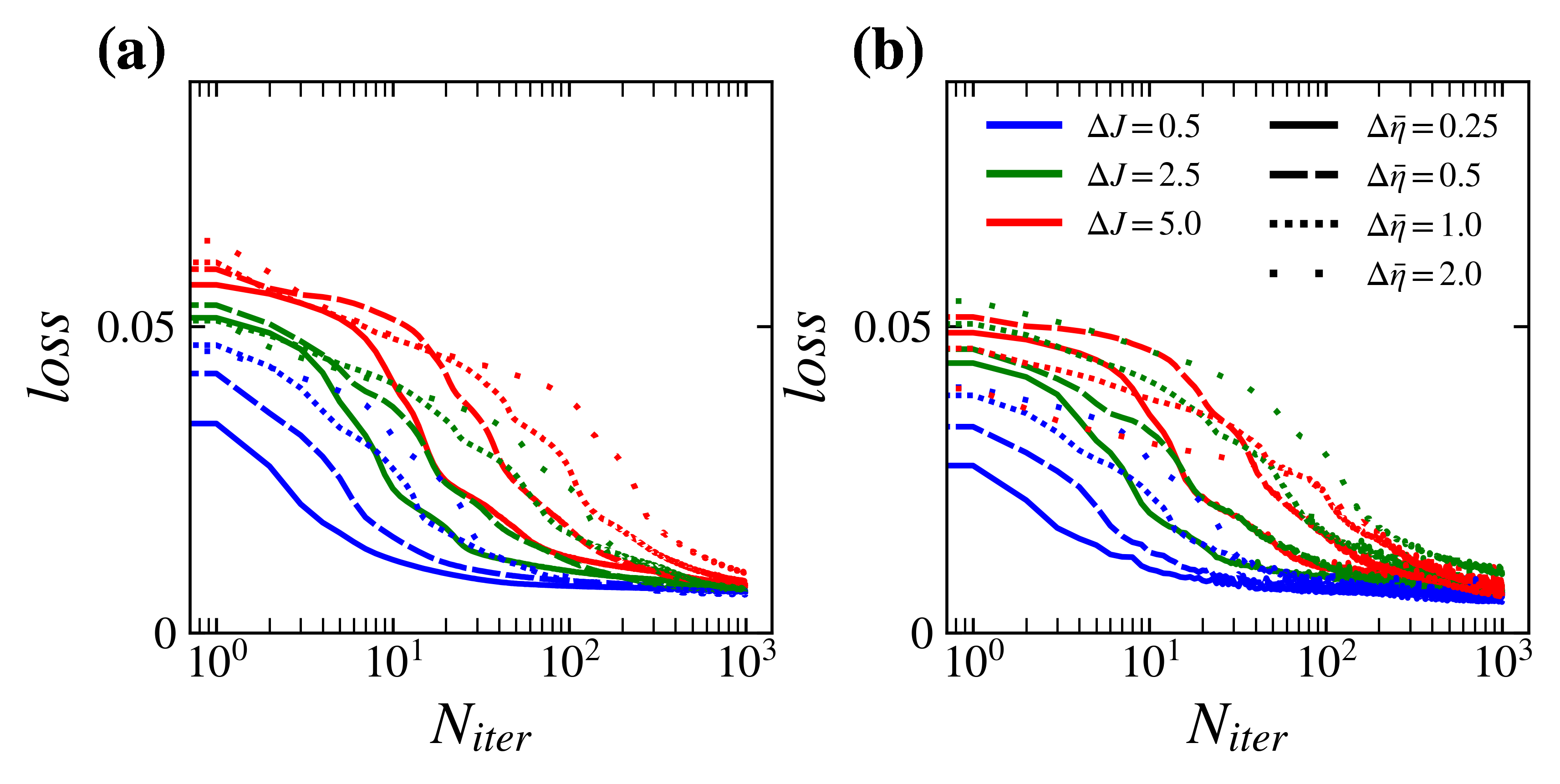}
\caption{\label{fig:loss} Loss curves for different datasets. (a) Training and (b) Test loss, for a $90/10,\  train/test$ split. Colors and line types indicate the sampling precision of $J$ and $\bar{\eta}$, respectively. In blue $\Delta J=0.5$, in green $\Delta J=2.5$, and in red $\Delta J=5$. The solid line corresponds to $\Delta \bar{\eta}=0.25$, the dashed line to $\Delta \bar{\eta}=0.5$, the dotted line to $\Delta \bar{\eta}=1$, and the loosely dotted line to $\Delta \bar{\eta}=2$.}
\end{figure}

\begin{figure}
\includegraphics[width=\columnwidth]{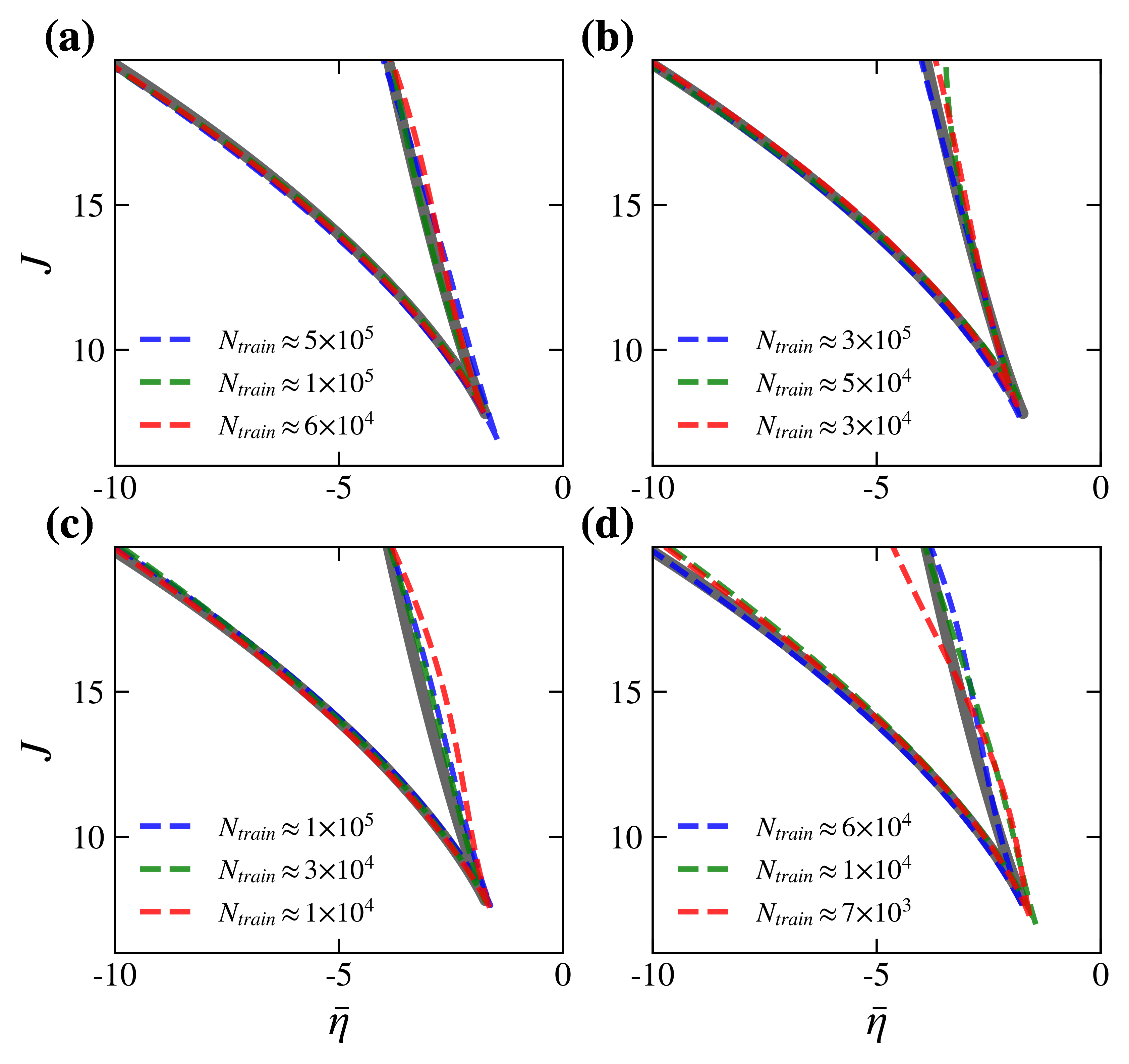}
\caption{\label{fig:bench_cusp} Quality of the reconstruction of the cusp bifurcation after $1000$ training iterations for different datasets. In (a) the sampling precision for $\bar{\eta}$ is $\Delta \bar{\eta}=0.25$, in (b) $\Delta \bar{\eta}=0.5$, in (c) $\Delta \bar{\eta}=1$ and in (d) $\Delta \bar{\eta}=2$. The line colors indicate the sampling precision for $J$, in blue $\Delta J=0.5$, in green $\Delta J=2.5$, and in red $\Delta J=5$. The shaded solid black in the four panels is the ground truth cusp from the $\text{MPR}$ model. The legend in the figure gives the approximate training dataset size for each combination of $\Delta J$ and $\Delta \bar{\eta}$.}
\end{figure}

To generate microscale data, we sampled $J$ and $\bar{\eta}$ on a regular grid. For validation purposes, we initially sampled the space with high density ($\Delta J=0.5$ and $\Delta \bar{\eta}=0.05$). In this section we provide a benchmark of the quality of the reconstruction of the dynamical features of the ground truth model for decreasing sampling density. We find satisfactory loss convergence in the training and test sets (\autoref{fig:loss}) even for coarse grained sampling. We reconstruct the 2-D cusp for different training set sizes (\autoref{fig:bench_cusp}) and find that even for coarse grained sampling in the parameter space ($\Delta J=5, \Delta \bar{\eta}=2$) the qualitative topology of the cusp is well retrieved. The lower bifurcation branch is systematically perfectly retrieve

\nocite{*}

\bibliography{references}
\end{document}